\documentclass[sigconf]{acmart}
\AtBeginDocument{%
  \providecommand\BibTeX{{%
    \normalfont B\kern-0.5em{\scshape i\kern-0.25em b}\kern-0.8em\TeX}}}

\usepackage{subfigure}

\definecolor{codegreen}{rgb}{0,0.6,0}
\definecolor{codegray}{rgb}{0.5,0.5,0.5}
\definecolor{codepurple}{rgb}{0.58,0,0.82}
\definecolor{backcolour}{rgb}{0.95,0.95,0.92}
\definecolor{bluepigment}{rgb}{0.2, 0.2, 0.6}

\newcommand{\greasedroid}{\textit{GreaseDroid}}
\newcommand\enquote[1]{``#1''}

\copyrightyear{2021}
\acmYear{2021}
\setcopyright{acmcopyright}\acmConference[CHI '21 Extended Abstracts]{CHI
Conference on Human Factors in Computing Systems Extended Abstracts}{May 8--13,
2021}{Yokohama, Japan}
\acmBooktitle{CHI Conference on Human Factors in Computing Systems Extended
Abstracts (CHI '21 Extended Abstracts), May 8--13, 2021, Yokohama, Japan}
\acmPrice{15.00}
\acmDOI{10.1145/3411763.3451632}
\acmISBN{978-1-4503-8095-9/21/05}

\let\oldttt\texttt
\renewcommand{\texttt}[1]{{\small \oldttt{#1}}}

\begin{document}
\title{I Want My App \textit{That} Way: Reclaiming Sovereignty Over Personal Devices}
\author{Konrad Kollnig}
\authornote{Both authors contributed equally to this research.}
\author{Siddhartha Datta}
\authornotemark[1]
\author{Max Van Kleek}
\email{{konrad.kollnig, siddhartha.datta,  emax}@cs.ox.ac.uk}
\affiliation{%
  \institution{Department of Computer Science, University of Oxford}
  \streetaddress{Parks Road}
  \city{Oxford}
  \postcode{OX1 3QD}
  \country{United Kingdom}
}

\renewcommand{\shortauthors}{Kollnig, Datta, and Van Kleek}

\begin{abstract}
Dark patterns in mobile apps take advantage of cognitive biases of end-users and can have detrimental effects on people's lives. Despite growing research in identifying remedies for dark patterns and established solutions for desktop browsers, there exists no established methodology to reduce dark patterns in mobile apps. Our work introduces \greasedroid{}, a community-driven app modification framework enabling non-expert users to disable dark patterns in apps selectively.
\end{abstract}



\maketitle

\begin{figure*}
    \centering
    \subfigure[Original Twitter app.]{\includegraphics[height=3in]{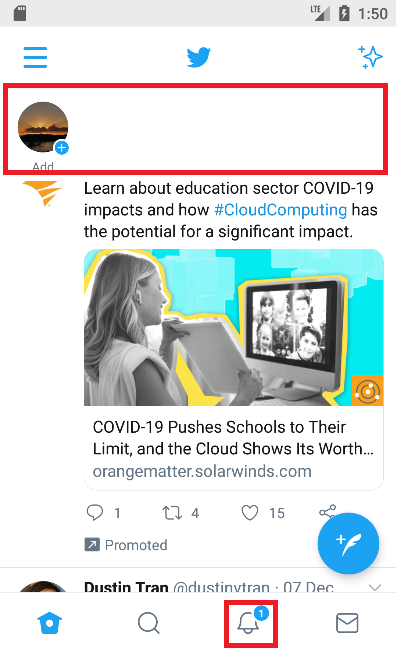}}
    \subfigure[\greasedroid{}-patched Twitter application with reduced dark patterns.]{\includegraphics[height=3in]{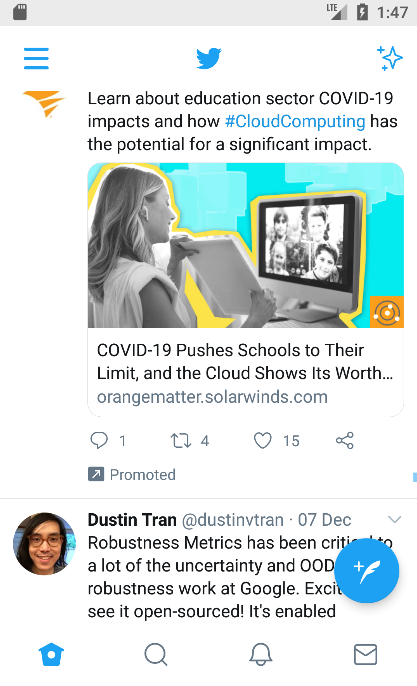}}
    \caption{
    \greasedroid{} enables the removal of dark patterns (highlighted in red) in Android apps.
    Compared to the default Twitter app (left), stories and notifications have been disabled to reduce distractions in the patched version of the app (right).
    }
    \Description{The figure shows a screenshot of the default Twitter app and a screenshot of the Twitter app with two dark patterns removed (namely notification highlights and Twitter stories).}
    \label{fig:ptch}
\end{figure*}

\section{Introduction}
  As our primary interfaces to both the digital world and to one
  another, mobile phones have become more than indispensable accessories, but extensions to our very beings that
  allow us operate as citizens within a distributed, global
  society~\cite{shklovski_leakiness_2014}. 
  Yet, the very mobile apps
  that now enable people to do so much are also being used as leverage
  by app developers, platforms, and services to exploit end-users, as
  evidenced by an increasing variety of user-hostile elements situated
  within services that provide this essential digital functionality~\cite{bundeskartellamt}.
  Such exploitation ranges from behavioural design (nudging people to
  take certain actions, such as making systems addictive), to
  exploiting users' psychological weaknesses, to features that
  directly harm the user such as by undermining their privacy either
  by giving up their personal information directly or by tracking them
  behind the scenes~\cite{lyngs_self-control_2019,lyngs_i_2020,binns_third_2018}.
  Even if apps do not pose direct harm to the majority of users, they often fail to account for the diversity and sensitivities of the most vulnerable groups of users~\cite{benjamin_race_2019}.
  Unfortunately, the tools for users to change and challenge their essential technological infrastructure are currently limited.
  Much of the essential software infrastructure is developed and deployed by powerful international tech companies, with limited means for user participation and negotiation in software design.
  What if users were give the ability to overcome user-hostile features in technology, thereby putting power back into the hands of its users?

This article explores how one specific example of user-hostile design in mobile apps, namely \textit{dark patterns}, can be alleviated.
Dark patterns are carefully-designed interfaces that induce a user to act with specific intent of the developer and widely used in mobile apps.
Studies that inspect UI elements on apps indicate high prevalence of dark patterns in web and software applications, typically covert, deceptive or information hiding in nature~\cite{10.1145/3359183, 10.1145/3313831.3376600, 10.1145/3290605.3300472}.
Dark patterns can have detrimental effects on data protection and consumer protection through the systematic exploitation of users' cognitive biases~\cite{10.1145/3319535.3354212, MultiplePurposesMultipleProblemsAUserStudyofConsentDialogsafterGDPR, 10.1145/3313831.3376321, soe2020circumvention}. This is particularly harmful to children, the aged, individuals with disabilities~\cite{wu_design_2019}, and the broader population \cite{rieger_sinders_2020}.
An intentionally-designed feature for one user may be considered a prohibitive bug for another.

This paper contributes to reducing the impact of dark patterns in mobile apps:
\begin{itemize}
    \item Our \emph{conceptual contribution} is a system for app modification as a way to reduce dark patterns in mobile apps. We believe that a community-driven approach can allow even non-expert users to modify apps in a way that fits their needs.
    \item Our \emph{technical contribution} is a working prototype of a user-scripting toolkit that manifests our general contribution to minimize dark patterns.
    We introduce sample patches with a case study of the Twitter app and show that dark patterns can be easily removed.
\end{itemize}

Whilst this article evaluates dark patterns, \greasedroid{} is by far not limited to this example of user-hostile design.
It can easily be used to modify other aspects of apps, including addictive design patterns and concerning privacy practices.

\section{Related Work}

As researchers in HCI, we strive to understand users and craft interfaces and systems for them. Unfortunately, user-focused research can be exploited against our users, with the design of interfaces that abuse cognitive and psychological vulnerabilities. It is proven that interfaces can inductively change a user's cognitive functioning~\cite{phoneabuse, LEE2014373, 10.5555/2821581}. Maliciously-crafted interfaces can thus induce precise cognitive functions intended by the developer, manifesting as digital addiction, digital persuasion, digital nudging, gamification, data exploitation, and dark patterns~\cite{10.1145/3334480.3381070}. 

This work focuses on \textit{dark patterns}, which can be defined as carefully-designed interfaces that induce a user to act with specific intent of the developer based on a user's cognitive biases.
\textit{Gray et al.} provide a 5-class taxonomy of dark patterns~\cite{10.1145/3173574.3174108}. 
\textit{(1) Interface Interference} elements manipulate the user interface such that certain actions are induced upon the user compared to other possible actions.
\textit{(2) Nagging} elements interrupt the user’s current task with out-of-focus tasks, usually in the form of a choice-based popup to redirect a user towards another goal.
\textit{(3) Forced Action} elements introduce sub-tasks forcefully before permitting a user to complete their desired task.
\textit{(4) Obstruction} elements introduce subtasks with the intention of dissuading a user from performing an operation in the desired mode.
\textit{(5) Sneaking} elements conceal or delay information relevant to the user in performing a task.

Given the prevalence of such interface patterns, more and more users wish to exercise \textit{digital self-control} to protect themselves. 
Despite the increasing body of literature on digital self-control, along with a wide array of anti-distraction apps, many users still struggle to exert meaningful control over their digital device use~\cite{lyngs_self-control_2019}.
It has been shown that interventions against harmful design patterns, such as hiding the Facebook feed in the desktop browser, can be effective at reducing usage time, and making users feel more in control of their device use~\cite{lyngs_i_2020}.
Yet, the limited ways for intervention on mobile devices are a major limitation of current digital self-control research and apps.

There is a long history of tools that try to fix deficiencies in the design of programs on \textit{desktops}, notably on websites.
One of these tools is the browser extension Greasemonkey. This tool lets users choose from a wide array of userscripts to change the design or functionality of a given website to fit their needs.
For instance, users can install userscripts to remove the feed from Facebook
or increase the readability of Wikipedia.
Similar technologies as used in Greasemonkey are the backbone of ad blockers, which are widely in use~\cite{stat}.

Greasemonkey and its variants can be considered examples of program repair tools. Program repair methods~\cite{martinez2016, ye2019automated, 10.1145/2786805.2786825} are concerned with the improvement of software and removal of bugs or errors after deployment. Dark patterns can pose functional hindrances, and could be considered as \enquote{bugs}. Dark patterns may leverage cognitive biases of the individuals that exceed the extent intended by or even beyond the intention of the developer, hence requiring certain dark patterns to be selectively mitigated. 

There exist some solutions for mobile devices to enhance the properties of apps. Methodologically, these solutions either 1) modify the operating system (e.g. Cydia Substrate~\cite{cydia}, Xposed Framework~\cite{xp}, ProtectMyPrivacy~\cite{agarwal_protectmyprivacy_2013}, or TaintDroid~\cite{enck_taintdroid_2010}), 2) modify apps (e.g. Lucky Patcher~\cite{lp}, apk-mitm~\cite{apkm}, Joel et al.~\cite{jeon_dr_2012}, DroidForce~\cite{rasthofer_droidforce_2014}, RetroSkeleton~\cite{davis_retroskeleton_2013}, AppGuard~\cite{garcia-alfaro_appguard_2014}, I-ARM-Droid~\cite{Davis12i-arm-droid:a}, Aurasium~\cite{Aurasium}), or 3) use System APIs (e.g. VPN-based ad blockers, such as AdGuard~\cite{ag} and TrackerControl~\cite{tcontrol}.

All of these solutions come with certain limitations. Whilst modifying the operating system can in principle make arbitrary modifications to the behaviour of apps, these usually rely on device vulnerabilities and are a moving target for device manufacturers. Operating system modifications can pose security risks, potentially void warranty, and are usually infeasible for non-expert users.
By contrast, the use of System APIs might often be the most straightforward approach for a non-expert user, operating in the familiar space of the user's own smartphone. This also poses a major limitation because only what is permitted by the smartphone operating system can be realised. For instance, the removal of the newsfeed from the Facebook app has not been accomplished through System APIs.

In app modification on Android, some transformation is applied to an app used by the user, e.g. the removal of the newsfeed from the Facebook app. App modification offers ease of use (installing custom apps on Android is supported by the operating system), and allows for arbitrary modifications of a provided app (only limited by the constraints of the operating system). As such, it combines benefits of system modification and System APIs.
Despite this, there exist hardly any solutions used in practice using app modification. One exception is the app cracking tool Lucky Patcher that allows to remove ads and paywalls from apps.
Another exception is \textit{apk-mitm} to remove certificate pinning from apps and allow to intercept encrypted network traffic.
Some developers have published modified versions of popular apps, including Facebook~\cite{xda} and YouTube~\cite{vanced} (removing ads and other distracting functions). Unfortunately, such modified apps rely on the continued support of the developers, may break over time (e.g. in case of server-side updates), and exist only for a few select apps.

\section{GreaseDroid}
\begin{figure*}
    \subfigure[General architecture.]{\includegraphics[height=1.9in]{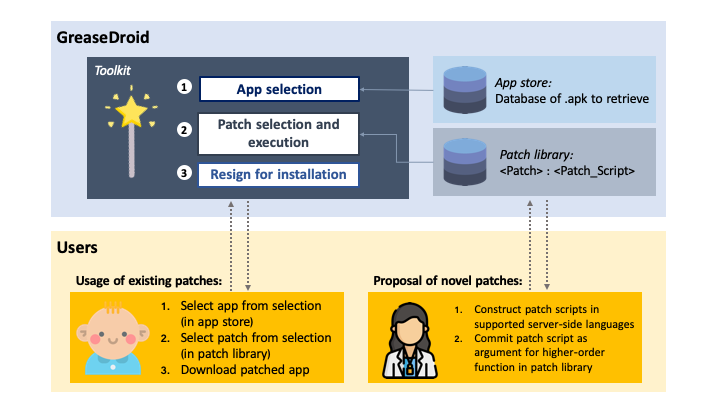}}
    \subfigure[Technical implementation.]{\includegraphics[height=1.9in]{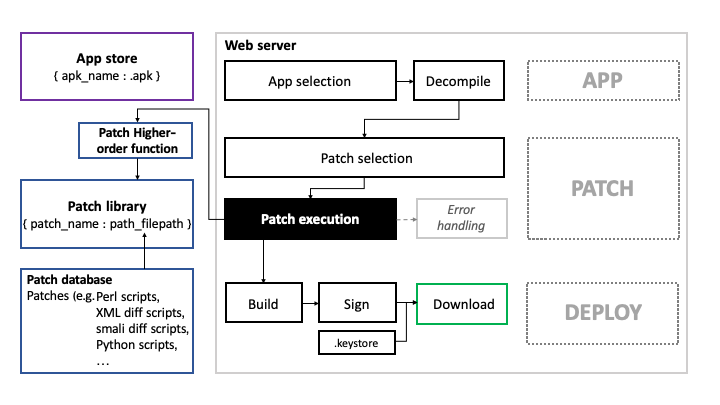}}
    \caption{Overview of the \greasedroid{} paradigm.}
    \Description{The figures show a schematic visualisation of the GreaseDroid approach. This is similar to the first figure that explains the three-step patching approach of our system.}
    \label{fig:archi}
\end{figure*}

To date, there exist no tool for non-expert smartphone users to remove dark patterns from their mobile apps.
We propose a patching paradigm based upon app modification, called \greasedroid{},
that enables user-scripting to reduce dark patterns in mobile apps.
\greasedroid{} acts as a tool that contains the assets needed for users to act autonomously in improving apps for their own wellbeing.
We retain a high level of abstraction for flexible implementation and additional modules; our implementation in code is covered in section \ref{imple}. Users go through three main phases (see in Figure~\ref{fig:archi}a). A user (1) selects an app, (2) applies a set of patches, and then (3) re-installs the apps on their phone. We share our code on GitHub\footnote{See {\color{blue}  \url{https://github.com/OxfordHCC/greasedroid}}.}.

\begin{figure}
    \centering
    \subfigure[Step 1: User selects an app.]{\includegraphics[width=3.2in]{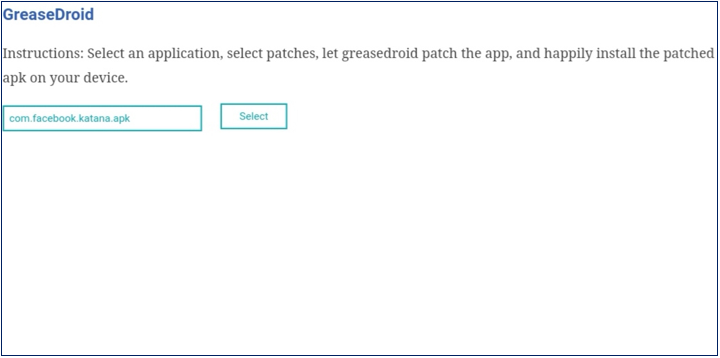}}
    \subfigure[Step 2: User selects patches.]{\includegraphics[width=3.2in]{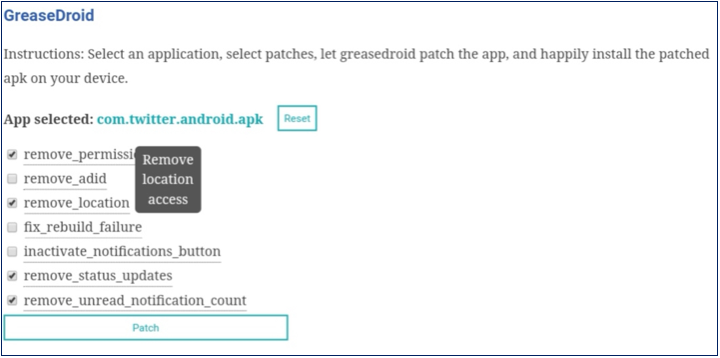}}
    \subfigure[Step 3: User installs patched app.]{\includegraphics[width=3.2in]{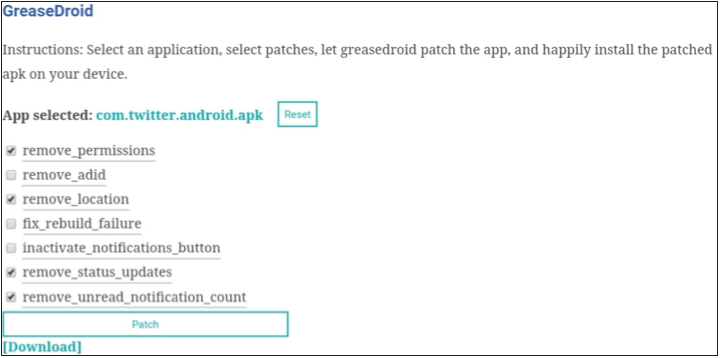}}
    \caption{Our implementation enables non-expert users to remove dark patterns from apps in three simple steps.}
    \label{fig:steps}
    \Description{The figure shows three screenshots of the three individual steps for app patching in the \greasedroid{} architecture.}
\end{figure}

\noindent\textbf{(1) App selection.}
In the first step, the user selects the app they want to patch.
\greasedroid{} shall allow the user to select arbitrary apps in the form of {\small \texttt{apk}} files (the standard format for apps on Android).
For increased ease of use, \greasedroid{} could directly fetch the latest version of the app from Google Play, or another third-party app store.
This would ensure that the user only ever patches the latest version of an app, and minimise unnecessary hassle of finding the {\small \texttt{apk}} file themselves.
Due to potential violations of Google's Terms \& Conditions, we have not yet implemented such an approach.

\noindent\textbf{(2) Patching.}
In the second phase, the user selects a set of patches 
to apply to the chosen app.
These patches are developed by expert community users, that we call \textit{patch developers}.
Once the user has chosen a set of patches, these are applied to the selected app and modify the app code (e.g. {\small \texttt{smali}} assembly code compiled libraries) or resources (e.g. {\small \texttt{xml}} layout files, images, sounds). 

Patches can be created in at least two ways, either as patch scripts or byte masks. Patch developers can create patch scripts by first decoding the app with {\small \texttt{apktool}}. They then sift through assembly code (stored as {\small \texttt{smali}} files and compiled libraries) and other resources (e.g. {\small \texttt{xml}} layout files, images, sounds) to identify potential code modifications for dark patterns. If the patches are generalizable over a number of apps, these are considered \textit{app-agnostic}, else \textit{app-specific}.
Our \greasedroid{} implementation uses patching based upon such patch scripts.
We will explain this method in more detail in the following sections.

Byte masks might be a powerful alternative to patch scripts.
These describe patches to apps that do not require decompilation, and instead directly modify the compiled bytecode of apps based on pattern matching.
For instance, colours within an app can be changed by simply searching for their hexadecimal representation in the app binary.
Similarly, the existing Lucky Patcher tool  replaces byte patterns in apps' compiled code to circumvent piracy protections~\cite{kannengiesser2016insight}.
Since byte masks do not rely on decompilation, they might be easier to deploy to a client-side patching system on the user's phone and might avoid legal issues related to decompilation.
We will not study this in more detail in this work because they are more difficult to develop for patch developers.

Ideally, patches crafted by patch developers should be designed to be executable on the patching device (e.g. a Linux-based patching environment may permit for a variety of patching scripts such as {\small \texttt{Perl}} or {\small \texttt{bash}}).
Patches should also take measures to be robust, including compatibility with patches and with updates to the app or operating system. 

\noindent\textbf{(3) Re-deployment.}
After successful app patching,
we need to re-sign in order to install the modified app on the user's device.
This re-signing process requires a user-generated certificate to ensure the authenticity of the app on the Android system.
It is important that the certificate is unique to the user, to prevent attackers from installing malicious apps on the user's phone.
A problem is that signing apps with user-generated certificate makes it impossible to install app updates directly from the Google Play store.
GreaseDroid offers an easy solution because the whole patching process can be repeated for any app updates.

\begin{figure*}
    \centering
    \includegraphics[width=\textwidth]{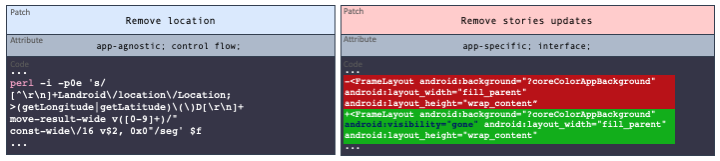}
    \caption{Left: An example of an \textit{app-agnostic Control Flow} patch -- a Perl script to prevent apps from accessing the longitude and latitude of the current physical user location. Right: An \textit{app-specific Interface} patch -- a diff script to remove the \textit{Stories} bar from the Twitter app.}
    \Description{GreaseDroid patches are modifications to the compiled source code of a mobile application. The figures show two examples of such modifications, that is, how the app code needs to be changed in order to incorporate two example patches.}
    \label{fig:code}
\end{figure*}

\section{Classes of Patches}

\greasedroid{} relies on the availability of patches.
To populate patches into the patch database, we rely on a network of expert patch developers to craft the patches.
Patches can be categorized based on their degree of app-specificity (app-agnostic or app-specific patches) and dark pattern blockage (interface or control flow patches).
These classes of patches are explained in the following.

\noindent\textbf{Control Flow patches vs Interface patches. } 
This category of patches is concerned with the types of dark patterns disabled in apps through patching. 
We refer to \textit{Gray et al.}'s taxonomy of dark patterns~\cite{10.1145/3173574.3174108}:
{\color{bluepigment} \textbf{Interface Interference}},
{\color{bluepigment} \textbf{Nagging}},
{\color{bluepigment} \textbf{Forced Action}},
{\color{bluepigment} \textbf{Obstruction}},
{\color{bluepigment} \textbf{Sneaking}}.
Each type of dark pattern can be tackled by either \textit{control flow patches} or \textit{interface patches}.
We expect most patches proposed to be interface patches, specifically targeting \textit{Interface Interference} dark patterns.
The distribution of dark patterns found in apps is skewed towards \textit{Interface Interference} elements, thus we expect the distribution of patching efforts to tend towards Interface Patches~\cite{10.1145/3313831.3376600, 10.1145/3170427.3188553}.

\textbf{Interface patches } [{\color{bluepigment} \textbf{Interface Interferences}}]
  Some of our patches directly change the UI of an app.
  Specific instances such as \textit{pre-selection} or \textit{hidden information} leverage the attention scope of end-users, so possible patches include the exaggeration of attributes of UI components to draw attention. Another instance, \textit{aesthetic manipulation}, might draw end-user attention to certain components, so possible patches include removing those components or aesthetically-modifying them to be less distracting.
  We show in Figure~\ref{fig:ptch} the concealment of elements that create distraction for users, such as the \textit{Stories} function and the notification counter. 
  Interface patches are often straightforward develop, requiring the modification of the Android {\small \texttt{xml}} layout files only; for example, changing visibility of elements, e.g. changing graphical asset colours to draw attention, rendering transparent or removing UI components.
  Such modification is similar to modifying {\small \texttt{html}} websites, and only requires expertise with app development on Android.

\textbf{Control flow patches } [{\color{bluepigment} \textbf{Nagging}},
{\color{bluepigment} \textbf{Forced Action}},
{\color{bluepigment} \textbf{Obstruction}},
{\color{bluepigment} \textbf{Sneaking}}]
  These patches require the modification of program control flow. 
  Patches in this category are more difficult to create. They rely on the modification of high-level assembly code in the {\small \texttt{smali}} programming language. It is challenging to modify the obfuscated assembly code, as integrity-preserving decompilation into {\small \texttt{Java}} for closed-source apps is not guaranteed. 
  Some dark patterns could be disabled by simply removing a function call to a pop-up function (for \textit{nagging}).
  Some dark patterns are built deep into the function of the app, and may require rewriting how functions execute (e.g. removing token mechanisms for \textit{gamefication}). 
  For example, \textit{forced actions} are in-task interruptions that can only be removed by modifying the control flow logic and refactoring the source code.

\noindent\textbf{App-agnostic vs. app-specific patches. }
This category of patches is concerned with range of apps that the patch can be applied to.
With our provided toolkit, we were able to craft patches for several apps, both (1) \textit{app-agnostic} patches (patches that can be applied to a wide range of applications) and (2) \textit{app-specific} patches (patches that are tailored to a specific application) (Figure \ref{fig:code}).

\section{Implementation}
\label{imple}

To test our proposed system of app patching to alleviate dark patterns, we constructed a prototype that enables end-users to apply pre-defined patches to Android apps of their choice. The following discusses this prototype implementation of our \greasedroid{} paradigm, consisting of end-users and patch developers.
Figure~\ref{fig:steps} shows screenshots of our implementation of the three-step patching process.
Figure~\ref{fig:archi}b illustrates our reference implementation of this flow in software.

End-users open the \greasedroid{} website on their device and select from a pre-selected list of Android apps to patch (Figure~\ref{fig:steps}a). They are shown a list of patches, then choose to apply their selected patches to the app (Figure~\ref{fig:steps}b). When complete, the user is provided a re-signed (Figure~\ref{fig:steps}c) installation file that can be installed on their phone.

Patch developers construct patches in the form of Linux scripts. These \textit{patch scripts} describe how to modify the code and resources of an app.
Each patch script comes with additional metadata, including name, description, author, and further robustness parameters (e.g. apps and app versions supported).
The use of Linux scripts for patching has multiple advantages.
First, it allows individuals with expertise in various programming languages to participate in user-scripting for \greasedroid{}.
Second,
Linux-based patches can potentially be run on the Android device itself, reducing the need for an external server and increasing the ease for the end-user.
For example, one of our app-agnostic patch scripts applies a regular expression through Perl to remove location access from an app, see Figure~\ref{fig:code}.

In our \greasedroid{} implementation, patching is executed on an external server. Patches can be created using common programming languages (e.g. {\texttt{python}}, {\texttt{Perl}}, {\texttt{bash}}) due to server pre-requisites. Patch developers insert a formatted patching higher-order function into the main patching library that calls and runs the patch script to patch a designated app. We provide interface patches that require modifications to {\texttt{xml}} layout and {\texttt{smali}} code files, so we first decode the {\small \texttt{apk}} file with {\texttt{apktool}}, then after calling the patch scripts, we run {\texttt{apktool}} to build a modified {\texttt{apk}} from source, and re-sign with user details. We then return a download link for user installation (Figure~\ref{fig:steps}c). To mitigate potential legal issues, we may consider switching from server-side to client-side patching on the user device in future work. 

\section{Case Study: Modification of the Twitter App}
\label{tweet}

To illustrate that \greasedroid{} can successfully remove dark patterns from apps, we consider the Twitter app on Android. In the main screen of the app, we identified two dark patterns that are used to encourage more user interaction, see Figure~\ref{fig:ptch}a. First, the top of the Twitter app shows so-called \enquote{Fleets}. These are tweets disappearing after 24 hours, similar to the \enquote{Stories} functionality of Facebook, Instagram and Snapchat. Second, the bottom of the app contains a notification counter that informs the users about recent activities of other users.
Neither Fleets nor notifications can be fully deactivated within the Twitter app, so some users might want to remove this functionality with \greasedroid{}, to reduce time spent on Twitter.

We were able to remove the two identified dark patterns from the Twitter app with the help of \greasedroid{}, by removing the relevant sections from the \texttt{xml} layout files from the decompiled Twitter app (i.e. from those files that describe the UI on Android).
It was straightforward to identify the relevant sections of the \texttt{xml} files with one of the run-time app layout inspection tools from the Google Play Store.
Additionally, with some more work, we injected code into the \texttt{smali} code of the decompiled Twitter app to refuse attempts to click the notification button, thereby disable access to the notification view of the Twitter app.

The study of the Twitter app demonstrates that it can be straightforward to modify UI components of existing apps.
Changing the program flow is more challenging than UI changes because the source code is usually obfuscated, but also possible for patch developers with good programming expertise.

\section{Discussion and Future Work}

\noindent\textbf{Benefits for app developers. } An ecosystem of patch developers and patch users is not isolated from the original app developers. Indeed, it may bring benefits for them.
The development of patches can serve as a feedback loop, to provide valuable suggestions for the original app developer.
With an active network of patch developers and patch adopters, \greasedroid{} can potentially speed up the app developer's development cycle
and reduce costs through crowdsourced software engineering efforts.
Indeed, the ability to create patches through \greasedroid{} might create new financial incentives, wherein app developers reward patch developers, similar to existing bug bounty programs.

An open question is how app developers will perceive increased user patching, whether it will be a moving target leading to an arms race, or whether developers will engage with the community so as to make apps better.

\noindent\textbf{Patch robustness across app updates. } App developers regularly release updates to their apps. This poses a particular challenge in designing patches because patches might lead to breakage of apps.
However, the amount of change across app updates will usually be rather small.
UI resources in Android apps are usually organised around \texttt{xml} layout files, which are easier to modify than compiled source code, as demonstrated by our case study of the Twitter app.
In addition, app-agnostic patches should not suffer from issues with app updates. A focus on UI and app-agnostic patches will help patch compatibility with app updates.

\noindent\textbf{Protection against malicious patches. } While community-driven user-scripting (1) reduces duplicate patch development and increases patch development efficiency, (2) increases patch quality selection (e.g. higher quality patches given a higher rating by patch adopters), and (3) increases collaborative opportunities within cliques in the patch developer network (e.g. identifying new patches, fixing bugs), a critical challenge is the threat of malicious patches. \textit{Greasemonkey}'s script market suffered from the existence of a significant number of malicious scripts disguised as benign scripts with abusable vulnerabilities~\cite{10.1145/2590296.2590311}. Though there have been recent developments in automated malware detection~\cite{10.1145/3134600.3134636, 10.1145/3134600.3134642, KARBAB2018S48, 8406618, 7846953, 8004891, 7814490}, due to the complexity and heterogeneity of patch scripts, automated approaches might not be sufficiently reliable to mitigate the threat of malicious patches~\cite{1e882f77dcb649b4aa4146efd6a07693, CHEN2018326, 7917369, 8782574}.
A review system of expert users might be preferable over automated program analysis to reduce malicious scripts. Each patch uploaded to the patch database could be \textit{verified} by a (group of) peer patch developer(s).
To ensure authenticity of reviews, patch reviewers could cryptographically sign patches after verification. 
The reliability of patch review by experts in practice may needed to be studied further.
Protecting users against malicious code is a great challenge in making \greasedroid{} available to a wide array of end-users, and will require more consideration and practical experience in future work.

\noindent\textbf{Legality. }
Using \greasedroid{} as an end-user might violate the law, particularly the DMCA in the US and the Computer Programs Directive in the EU.
Two main issues arise: 1) the distribution of patched apps and 2) the decompilation and disassembly of apps.
The decentralised approach of \greasedroid{} might help to overcome such legal challenges. Whilst developers previously shared patched apps online, \greasedroid{} separates the distribution of patches from the patching of apps.
Patches are applied at install-time and on the user's device. There is no need to distribute patched apps.
Such private modification of apps might be covered by the \enquote{fair use} and \enquote{right to repair} principles in the US because \greasedroid{} allows users to remove deficiencies from apps.

Our implementation of patching in \greasedroid{} does not rely on \textit{decompilation} of the program code, but rather on \textit{disassembly}. This could reduce issues with legislation that bans decompilation. Byte masks -- as used in LuckyPatcher -- could even remove the need for disassembly, and instead make changes to the app binary directly.
Under the EU Computer Programs Directive, decompilation is permitted when \textit{\enquote{necessary for the use of the computer program by the lawful acquirer in accordance with its intended purpose, including for error correction}}.
Enabling disadvantaged users to use essential technologies; promoting research in the public interests; and complying with fundamental human rights laws, including privacy and democratic rights, may well be the intended purpose of many apps.

\section{Conclusion }

This work describes \greasedroid{}, a framework to reduce dark patterns in mobile apps through app modification, and provide a functional prototype implementation. Using the example of the Twitter app, we illustrate the patch development process and prove the functionality of our method.
The successful deployment of community-driven user-scripting to minimize dark patterns in mobile apps can be further pursued with research in digital self-control, app privacy, software law, software engineering practices, and app security and reverse-engineering. Further work should study how our app modification framework can be deployed more widely, particularly mitigating the technical and legal challenges; why users would want to modify their apps; how effective app modification is at helping users gain more agency in their app use. This work aims to open the debate around what choice users should have over their apps, and whether there should be a \enquote{right to fair programs}, even if this requires changes to existing legislation.
Given that users rely on their apps and have currently limited ways to change them, \greasedroid{} introduces means to change their apps, and negotiate the terms of their apps.


\bibliographystyle{ACM-Reference-Format}
\bibliography{bib}

\end{document}